\documentclass[a4paper,11pt]{article}
\usepackage{pos}

\title{Calculating the energy loss of leading jets}

\author[a]{Duff Neill}
\author[b]{Felix Ringer}
\author[c]{Nobuo Sato}

\affiliation[a]{Theoretical Division, MS B283, Los Alamos National Laboratory, Los Alamos, NM 87545, USA}

\affiliation[b]{Nuclear Science Division, Lawrence Berkeley National Laboratory, Berkeley, California 94720, USA}

\affiliation[c]{Theory Center, Jefferson Lab, Newport News, Virginia 23606, USA}

\emailAdd{duff.neill@gmail.com}
\emailAdd{fmringer@lbl.gov}
\emailAdd{nsato@jlab.org}

\abstract{The energy loss mechanism of jets plays a central role in nuclear and high energy physics. We propose direct measurements of the energy loss of leading jets and perform a calculation at next-to-leading logarithmic (NLL$'$) accuracy in the vacuum. The formation of leading jets can be described by jet functions which constitute probability densities and thus allow for a perturbative calculation of the average the energy loss. We identify the following three criteria for a direct measurement of jet energy loss at the cross section level. $i)$ We measure a well defined object, the leading jet, where the formation process can be expressed in terms of a probability density. $ii)$ In addition, we need a measurement of a hard reference scale with respect to which jet energy loss is defined. $iii)$ At leading logarithmic accuracy, we require that the jet energy loss can be identified with parton energy loss. We discuss suitable observables and present numerical results including threshold corrections by making use of a parton shower Monte Carlo approach.}

\FullConference{%
  HardProbes2020\\
  1-6 June 2020\\
  Austin, Texas}

\begin{document}
\maketitle

\section{Introduction}

\begin{figure}[t]
\vspace*{.7cm}
\centering
\includegraphics[width=12cm]{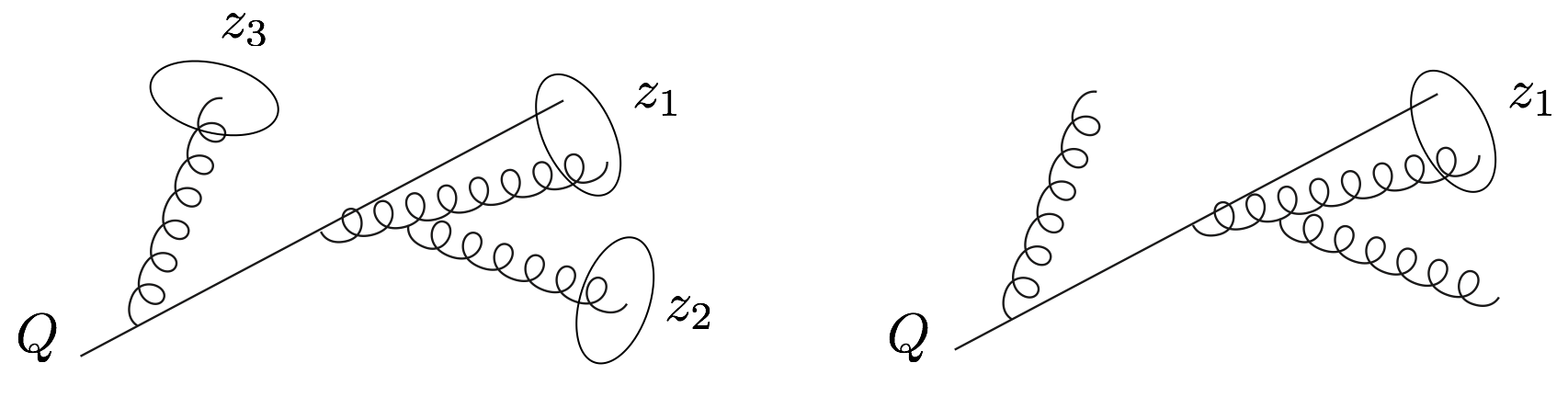}
\caption{Illustration of inclusive jets (left) and the leading jet (right) originating from a fragmenting quark. We indicate the momentum fractions $z_i$ of the jets relative to the initial scale $Q$.~\label{fig:inclusive_leading}}
\end{figure}

In heavy-ion collisions the jet energy loss mechanism due to medium-induced emissions plays a crucial role in quantitatively understanding the formation and properties of the quark-gluon plasma~\cite{Gyulassy:1993hr,Baier:1996sk,Zakharov:1996fv,Gyulassy:2000er,Arnold:2002ja,CasalderreySolana:2011us}. Here we propose a direct measurement of the energy loss of leading jets and we perform a calculation at next-to-leading logarithmic (NLL$'$) accuracy in the vacuum. Similar to inclusive jets~\cite{Catani:2013oma,Dasgupta:2014yra,Kaufmann:2015hma,Kang:2016mcy,Dai:2016hzf}, the factorization of leading jets can be written in terms of hard-scattering functions and jet functions. As illustrated in Fig.~\ref{fig:inclusive_leading} (right), the leading jet function ${\cal J}_i$ takes into account the formation of the jet which carries the largest momentum fraction $z_1$ relative to the initial scale $Q$. It constitutes a probability density for finding the leading jet which can be determined perturbatively. By calculating the mean of the jet function, or a suitably defined cross section, we can quantify the average energy fraction contained in the leading jet $\langle z_{i,1}\rangle$ for $i=q,g$. The energy outside the leading jet, the jet energy loss, is thus given by $\langle z_{i,{\rm loss}}\rangle=1-\langle z_{i,1}\rangle$. An inclusive jet sample (Fig.~\ref{fig:inclusive_leading}, left) is obtained by taking into account all jets independent of their momentum fraction $z_i$ and instead of energy loss, we measure a redistribution of energy through the branching process. We denote corresponding inclusive jet function by $J_i$. We identify the following three criteria which allow for a direct measurement of the jet energy loss at the cross section level
\begin{itemize}
    \item We measure a well defined object, the leading jet, which has lost energy relative to an initial scale where the formation process is described by a probability density. We note that this does not apply to inclusive jets where the corresponding jet functions $J_i$ are number densities.
    
    \item In addition, in order to quantify jet energy loss we need to measure a hard reference scale with respect to which we define the energy loss of the leading jet. We discuss possible observables below.
    
    \item We require that the (average) jet energy loss can be identified with the energy loss of a fragmenting quark or gluon at leading logarithmic accuracy.
\end{itemize}

\noindent
The leading jet functions satisfy non-linear evolution equations~\cite{Waalewijn:2012sv,Dasgupta:2014yra,Elder:2017bkd,Scott:2019wlk}. We employ a parton shower Monte Carlo approach to solve these equations and we include threshold corrections at next-to-leading logarithmic (NLL$'$) accuracy. In addition, the threshold resummed hard functions which appear in the factorization formulas of leading jet cross sections are also directly included in the parton shower algorithm.

\section{Theoretical framework}

We compare next-to-leading order (NLO) results, evolution equations and the factorization of inclusive and leading jets. The cross section of inclusive jets can be calculated by employing a factorization in terms of hard and jet functions $J_i$ which was introduced in~\cite{Catani:2013oma,Dasgupta:2014yra,Kaufmann:2015hma,Kang:2016mcy,Dai:2016hzf}. The jet functions satisfy DGLAP evolution equations which allow for the resummation of logarithms of the jet radius $R$. At NLO, we have only two partons which are clustered into a single jet or two separate jets. The NLO leading jet functions ${\cal J}_i$ can be expressed in terms of the inclusive ones as
\begin{equation}
    {\cal J}_{i}(z_1,Q R,\mu)\stackrel{{\rm NLO}}{=} \Theta(z_1>1/2)\,J_i(z_1,Q R,\mu)\,.
\end{equation}
Here we write the jet functions in terms of the hard scale which is set for example by the initial transverse momentum $Q=\hat p_T=p_T/z_1$ of the fragmenting quark or gluon. The theta function indicates that we only consider the parton or jet with the larger momentum fraction in the case that the two partons are clustered into separate jets. The spectrum in $z_1$ of the leading jet function extends down to $z_1>1/(n+1)$ at order $n$ in perturbation theory, though the physical spectrum is for $0<z_1<1$. Therefore, the all order resummation discussed below is critical to access the full spectrum. An important difference between inclusive and leading jet functions is that they constitute number densities and probability densities, respectively. We have therefore the sum rules:
\begin{align}
    \int_0^1{\rm d}z \, J_i(z,Q R,\mu)=\,\langle N_{i,{\rm jets}} \rangle 
    \,,\quad\quad 
    \int_0^1{\rm d}z_1 \, {\cal J}_i(z_1,Q R,\mu)=\,1\,,\label{eq:normalization}
\end{align}
where $\langle N_{i,{\rm jet}}\rangle$ denotes the average number of jets originating from a quark or gluon. This average number is not conserved, by is dynamically generated in the collisions. From the leading jet function we can thus calculate the average momentum fraction contained in the leading jet
\begin{equation}\label{eq:eloss}
    \int_0^1 {\rm d}z_{1}\, z_{1} \, {\cal J}_{i}(z_{1},Q R,\mu)=\langle z_{i1}\rangle \,,
\end{equation}
and the energy loss is given by $\langle z_{i,{\rm loss}}\rangle = 1-\langle z_{i,1}\rangle$. Note that for inclusive jets this integral is equal to unity due to momentum conservation. For example, for anti-k$_T$ jets~\cite{Cacciari:2008gp} we find that the energy loss of a quark and gluon at NLO is given by
\begin{align}
    \langle z^{k_T}_{q,{\rm loss}} \rangle  =&\, \frac{\alpha_s}{2\pi}C_F \ln(1/R^2)\left(2\ln 2-\frac38 \right)+\frac{\alpha_s}{2\pi}C_F\left(4\ln^2 2+\frac32 \ln 2 - \frac{19}{8}+ \frac{\pi^2}{3}\right)
    \,,\\
    \langle z^{k_T}_{g,{\rm loss}} \rangle 
    =&\, 
    \frac{\alpha_s}{2\pi}\ln(1/R^2)\left[C_A \left(2 \ln 2-\frac{43}{96} \right)+ N_f T_F \frac{7}{48}  \right]
    \nonumber\\&
    +\frac{\alpha_s}{2\pi}\left[C_A \left(4 \ln^2 2 + \frac{15}{8} \ln 2 -\frac{793}{288} + \frac{\pi^2}{3}\right) +N_f T_F \left(- \frac{3}{4} \ln 2 +\frac{65}{72}\right)\right] \,.
\end{align}
In order to include higher order corrections, we need to solve the DGLAP type evolution equations of the leading jet functions which are non-linear
\begin{align}
    \mu\frac{{\rm d}}{{\rm d}\mu}{\cal J}_{i}(z_{1i},Q R,\mu)=\,&\frac12 \sum_{jk}\int {\rm d}z\, {\rm d}z_{j1} {\rm d}z_{k1} \frac{\alpha_s(\mu)}{\pi}P_{i\to jk}(z)\, {\cal J}_{j}(z_{j1},Q R,\mu)\, {\cal J}_{k}(z_{k1},Q R,\mu) 
    \nonumber\\&\times 
    \delta(z_{i1}-\text{max}\{z z_{j1},(1-z)z_{k1}\})\,,
\end{align}
and evolve the jet functions between the scales $QR$ and $Q$. Note that the normalization in Eq.~(\ref{eq:normalization}), and thus the probabilistic interpretation, is conserved under the evolution. The non-linear evolution equations can be solved iteratively or by means of a parton shower Monte Carlo approach which we choose here. We extend the work of~\cite{Dasgupta:2014yra,Scott:2019wlk} by evolving the full NLO jet function and by including threshold corrections at NLL$'$. The threshold resummation is carried out analytically following Refs.~\cite{Becher:2006mr,Dai:2017dpc,Liu:2017pbb}. We note that in the threshold limit considered below, the resummed hard function can be convolved directly with the evolved leading jet functions, e.g. ${\cal H}_i\otimes {\cal J}_i(z_1)$. We include the threshold resummed hard and jet functions directly in the parton shower cascade, and the measurement of the leading jet is carried out at the very end.  More details of the parton shower approach employed here will be presented in Ref.~\cite{forwardcite}.

\begin{figure}[t]
\vspace*{.7cm}
\centering
\includegraphics[width=\textwidth]{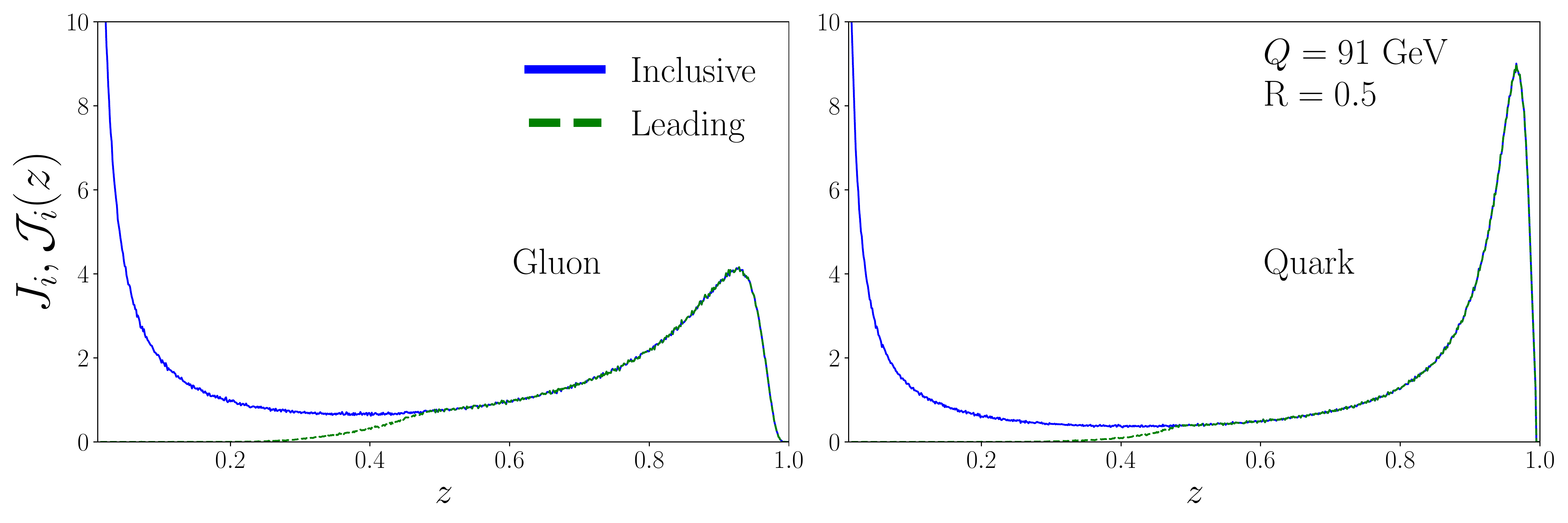}
\caption{The inclusive and leading jet fragmentation spectrum with $R=0.5$ and hard scale $Q=91$~GeV for gluons (left) and quarks (right).~\label{fig:inclusive_leading_NLL}}
\end{figure}

\section{Numerical results}

Following the criteria listed in the Introduction, we find that the following processes are suitable for a direct measurement of the leading jet energy loss. In proton-proton or heavy-ion collisions, the reference scale can be set by a photon recoiling the leading jet~\cite{Sirunyan:2017qhf,Aaboud:2018anc,Acharya:2020sxs}. Alternatively, we can construct a reference scale by first measuring jets with jet radius $R$ and transverse momentum $p_T$ and we then determine the energy spectrum of leading subjets with $r<R$ relative to $Q=p_T$. In addition, high energy collisions with initial state leptons allow for alternative reference scales. In Semi-Inclusive Deep Inelastic Scattering~\cite{Arratia:2020ssx} and in $e^+e^-$ collisions, a hard reference scale is set by the photon virtuality $Q^2=-q^2$ and the center-of-mass energy $Q=\sqrt{s}$, respectively. For the different processes listed here only the (threshold resummed) hard function needs to be changed. For illustration purposes we consider in this work $e^+e^-$ collisions with a $q\bar q$ and $gg$ final state and we identify the leading and inclusive jets in one hemisphere. The resulting inclusive and leading jet spectra are shown in Fig.~\ref{fig:inclusive_leading_NLL} for $R=0.5$ and a hard scale of $Q=91$~GeV for quarks and gluons. For $z>1/2$ the two spectra agree since in this regime there is only one jet which is automatically the leading one. We observe a peak close to one which indicates that it is likely to find a jet that carries a large fraction of the initial momentum. This behavior of the spectrum is very different compared to the fragmentation process of a hadron. The peak structure of the perturbative spectrum is due to the threshold resummation. As expected the peak is more smeared out for gluons than for quarks. Below $z=1/2$, the inclusive spectrum differs from the leading jet result and rises toward small-$z$. The leading spectrum falls off as it is unlikely to find a leading jet which carries only a small momentum fraction. The difference between the two results is given by the ($n$-th) subleading jets which start contributing below $z=1/(n+1)$. We note that the transition at $z=1/2$ is smoother after including higher order corrections.

\begin{figure}[t]
\vspace*{.7cm}
\centering
\includegraphics[width=.47\textwidth]{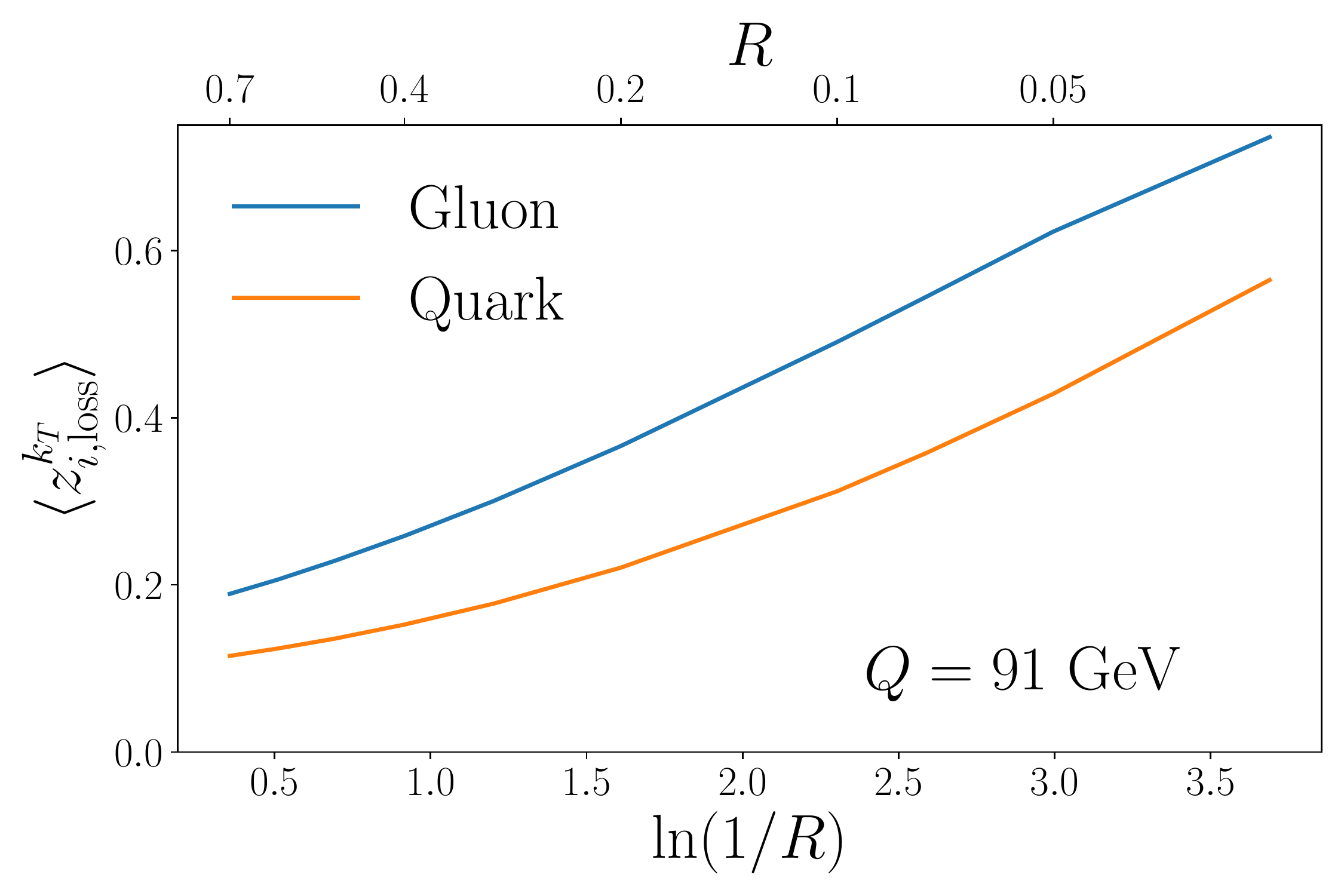}
\hfill
\includegraphics[width=.47\textwidth]{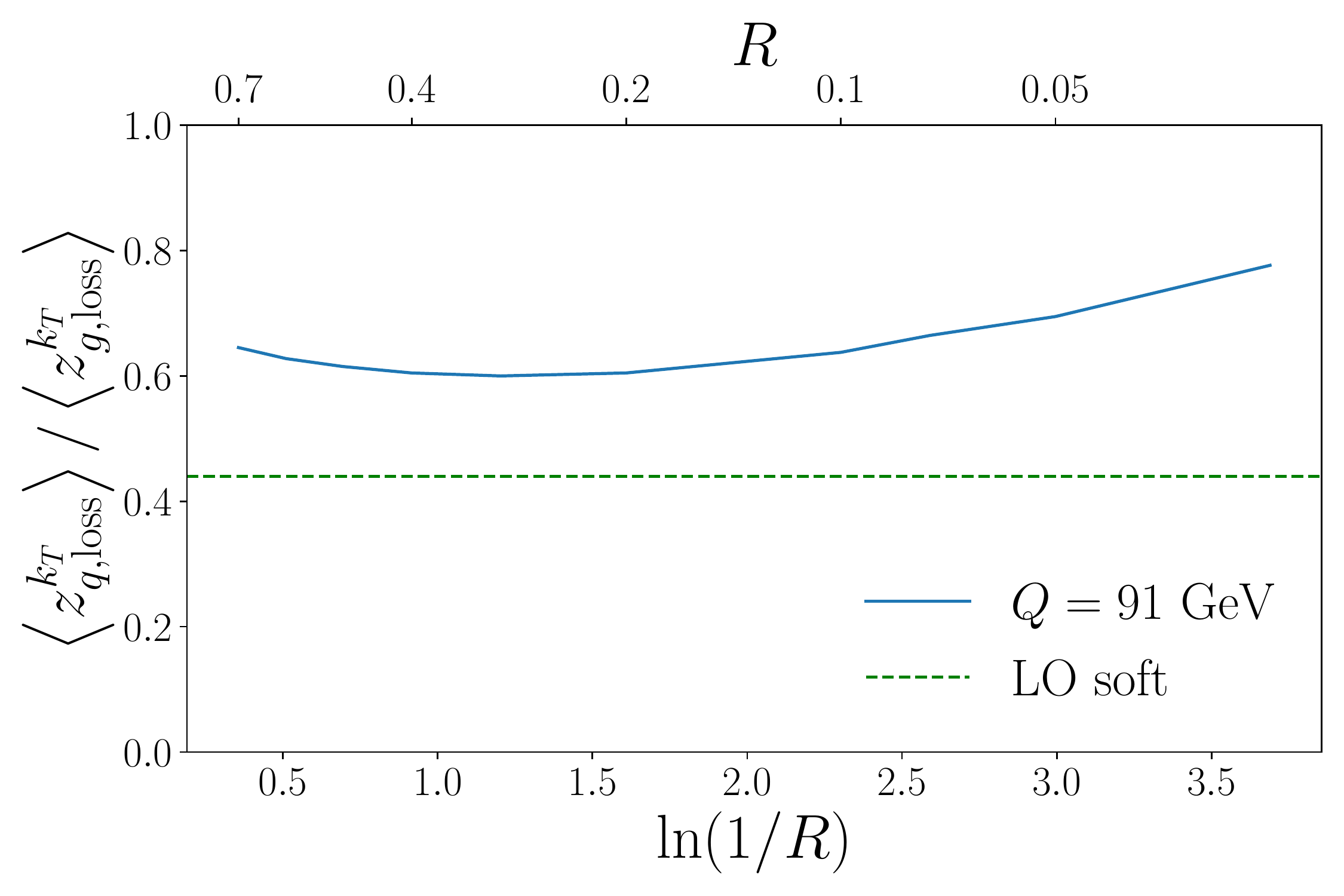}
\caption{The average energy loss of a leading quark and gluon jet (left) and their ratio (right) as a function of the jet radius for a hard scale of $Q=91$~GeV.~\label{fig:eloss}}
\end{figure}

Next, we consider the average energy loss which is obtained from the leading jet spectra above by taking the first moment, see Eq.~(\ref{eq:eloss}). In Fig.~\ref{fig:eloss} (left), we show the average energy loss of a quark and gluon jet as a function of the jet radius $R$. As expected the energy loss of the leading jet grows as $R$ is decreased as less radiation can be captured inside a narrow jet. In addition, the energy loss of gluons is larger than for quarks. To make this more clear, we plot the ratio of the quark and gluon jet energy loss which is shown in the right panel of Fig.~\ref{fig:eloss}. The difference between quarks and gluons turns out to be largely independent of the jet radius $R$. The dotted line shows a leading-order estimate of the difference between quarks and gluons which, in the soft limit, is given only by a ratio of color factors $C_F/C_A\approx 0.44$. After including the higher order corrections discussed above, we find that the ratio is actually significantly higher $\approx 0.6-0.7$. We conclude that quark/gluon differences of the jet energy loss are less pronounced than expected from a leading-order estimate.

\section{Conclusions}

In this work, we proposed observables to directly measure the energy loss of the leading jet. The formation of leading jets is described by non-linear DGLAP type evolution equations. We performed a (vacuum) calculation of the jet energy loss distribution and its average at next-to-leading logarithmic (NLL$'$) accuracy. We developed a parton shower approach where we included also the threshold resummation of both the hard and the jet function. We presented results for the average energy loss and found that the difference between quark and gluon jets is smaller than expected compared to a simple leading-order argument based on color factors. We expect that our work and future experimental measurements can shed new light on the energy loss mechanism in the vacuum and in heavy-ion collisions as well as electron-nucleus collisions at the future Electron-Ion Collider. Further details of our work will be presented in Ref.~\cite{forwardcite}.

\section*{Acknowledgements}

We thank Miguel Arratia, Peter Jacobs, Yen-Jie Lee, Yiannis Makris, James Mulligan, Dennis Perepelitsa, Mateusz Ploskon, Darren Scott and Wouter Waalewijn for helpful discussions. D.N. was supported by the U.S. DOE under Contract DE-AC52-06NA25396 at LANL and through the LANL/LDRD Program. F.R. was supported by LDRD funding from Berkeley Lab provided by the U.S. Department of Energy under Contract No. DE-AC02-05CH11231. N.S was supported through DOE Contract No. DE-AC05-06OR23177 under which JSA operates the Thomas Jefferson National Accelerator Facility

\bibliographystyle{JHEP}
\bibliography{bibliography}

\end{document}